# Reflecting on a Decade of Formalized Tornado Emergencies

Edward C. Wolff,[a] James S. Goodnight,[b] Leanne Blind-Doskocil,[c] Elijah M. Conklin,[c] Evan T. Gustafson,[c] and Joseph E. Trujillo-Falcón[a,d]

[a] *Department of Climate, Meteorology, & Atmospheric Sciences*, *University of Illinois Urbana-Champaign, Urbana, IL*

[b] *Lawrence Livermore National Laboratory, Livermore, CA*

[c] *Department of Geography, Meteorology, and Environmental Science, Valparaiso University, Valparaiso, IN*

[d] *Department of Communication*, *University of Illinois Urbana-Champaign, Urbana, IL*

*Corresponding author*: Edward C. Wolff, ecwolff3@illinois.edu

ABSTRACT

In 1999 the NWS began using the phrase "tornado emergency" to denote tornado warnings for storms with the potential to cause rare, catastrophic damage. After years of informal usage, tornado emergencies were formally introduced to 46 weather forecasting offices in 2014 as part of the impact-based warning (IBW) program, with a nationwide rollout occurring over the following years. In concert with the new tiered warning approach, the Warning Decision Training Division (WDTD) also introduced suggested criteria for when forecasters should consider upgrading a tornado warning to a tornado emergency, which includes thresholds of rotational velocity (VROT) and significant tornado parameter (STP). Although significant research has studied both tornado forecasting and tornado warning dissemination in the decade since, relatively little work has examined the effectiveness of the tornado emergency specifically. Our analysis of all 89 IBW tornado emergencies issued from 2014-2023 found that forecasters do not appear to follow the suggested criteria for issuance in the majority of cases, with only a handful of tornado emergencies meeting both the VROT and STP thresholds. Regardless, 70% of tornado emergencies were issued for EF-3+ tornadoes, and tornado emergencies covered 55% of all EF-4 tornadoes as well as 41% of all tornadoes resulting in 3 or more fatalities. Based on these results, we propose several updates to the current NWS training materials for impact-based tornado warnings.

SIGNIFICANCE STATEMENT

This article examines the first decade of formalized tornado emergencies, specifically how closely forecasters adhere to guidelines for issuance and whether emergencies successfully identify the most dangerous tornadoes. We find that forecasters rarely consider all of the recommended criteria and may instead be developing practices tailored to their unique regional contexts. Nevertheless, the majority of violent tornadoes have been warned with tornado emergencies, while false alarms have been minimal. We conclude with practical recommendations to provide a pathway toward improved operational guidelines grounded in current forecasting practice.

CAPSULE

An analysis of U.S. tornado emergencies from 2014–2023 reveals forecasters more often meet thresholds that are radar-based than environment-based and may also consider tornado width and population density.

## 1. Introduction

Each year, hundreds of tornadoes occur in the United States. While most have maximum winds ≤ 135 mph (i.e., rated EF-2 or lower on the enhanced Fujita scale), a handful of tornadoes contain intense winds, capable of destroying well-built structures. These especially intense tornadoes, rated EF-3 or higher, made up less than 2.5% of all tornadoes from 2014-2023 but accounted for 76.5% of fatalities according to an analysis of the NCEI Storm Events Database (NCEI 2025). The clear distinction in severity of impacts between weak and intense tornadoes would seem to necessitate a similar distinction in the warnings issued for these storms by the NWS. Such a distinction has been attempted in the past decade through the formalization of the term "tornado emergency."

The phrase tornado emergency has been used informally since 1999 (Andra et al. 2002) but was only formally introduced to a subset of NWS Weather Forecasting Offices (WFOs) in 2014 with the advent of the impact-based warning (IBW) program (Hudson et al. 2013; 2015). In the years since, IBWs have been integrated into all WFOs. The IBW framework takes advantage of advancements in nowcasting tornado intensity, such as additional research on near-storm environments and radar-based storm characteristics (Hudson et al. 2015). This framework consists of 3 levels or "tags" for tornado warnings: base, considerable, and catastrophic (NWS 2023). The "considerable" tag is colloquially referred to as a particularly dangerous situation (PDS) tornado warning while the "catastrophic" tag corresponds to a tornado emergency (NWS FFC 2015), with this highest level being used in "exceedingly rare" situations where reliable sources have confirmed a tornado capable of extreme damage (NWS 2023).

To aid forecasters, the Warning Decision Training Division (WDTD) created a set of scientific recommendations for when forecasters should consider using each of the new warning "tags" (Warning Decision Training Division 2023a; Fig. 1). Forecasters are recommended to consider upgrading a tornado warning to an emergency if they observe rotational velocity (VROT) ≥ 70 kt on radar and significant tornado parameter (STP) ≥ 6 in the near-storm environment.[1] These two criteria should also be accompanied by either radar or spotter confirmation of an ongoing tornado, with the former being given by the presence of a tornado debris signature (TDS; Ryzhkov et al. 2005; Van Den Broeke and Jauernic 2014).

[1] Personal communication with WDTD staff has indicated that the specific criteria shown in Fig. 1 were formally introduced in December 2022 and that criteria are consistently evolving as new studies are published and operational capabilities change.

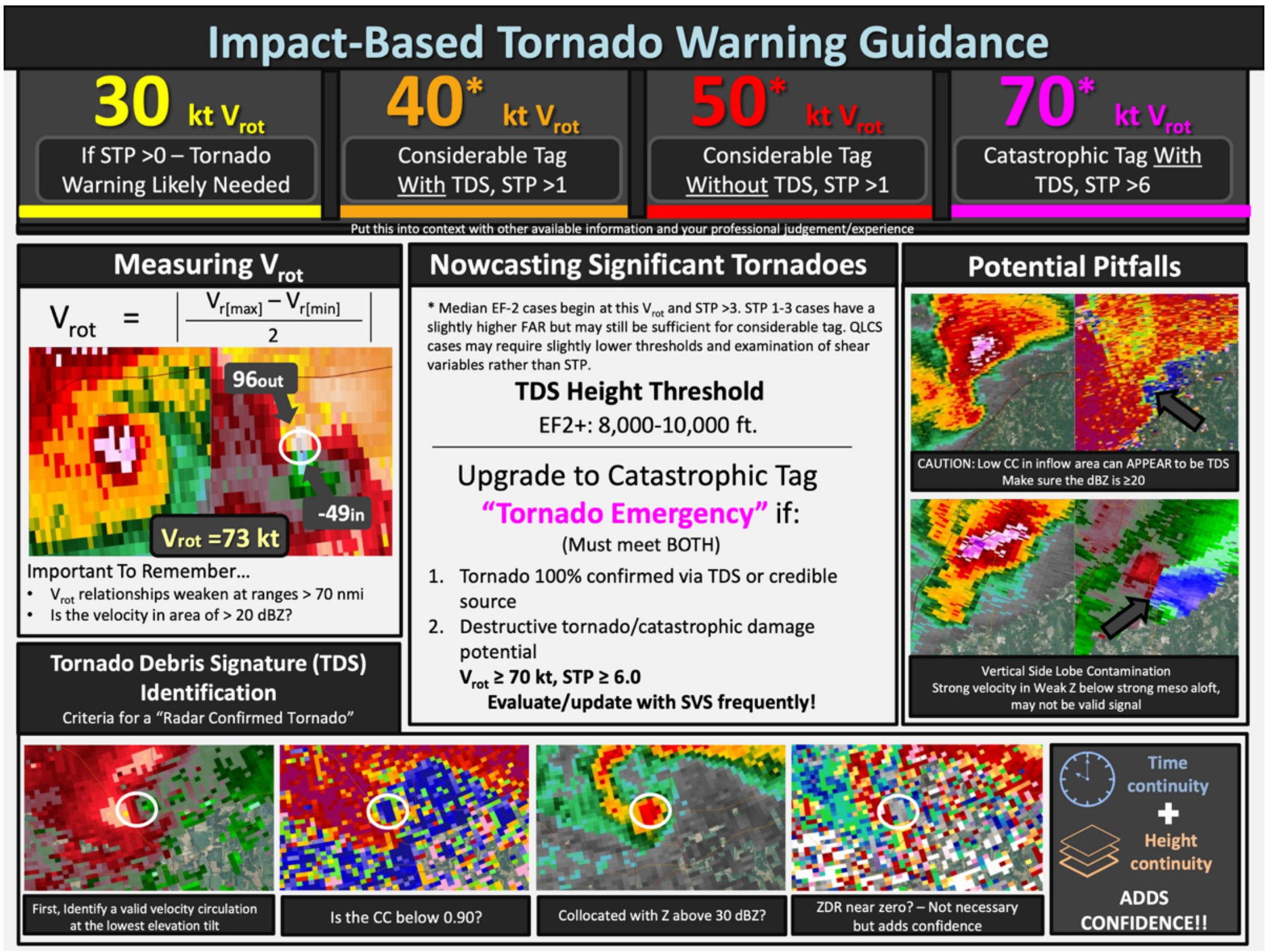

**Fig. 1.** Reference guide created by the NWS WDTD with guidance for issuing IBWs, including the suggested criteria for upgrading a tornado warning to a tornado emergency: tornado confirmation via a TDS or credible source and catastrophic damage potential, as indicated by VROT ≥ 70 knots and STP ≥ 6 (Courtesy of the NWS WDTD).

These thresholds are rooted in climatologies of supercell tornadoes, such as Smith et al. (2015) and Bentley et al. (2021), the former of whom showed both VROT and STP were strongly related to peak tornado intensity, though there was considerable overlap in STP values between EF categories. Older training modules from the WDTD identify population density as another important criterion, which originated from the threat of tornadoes in major metropolitan areas (Dunn and Vasiloff 2001; Wurman et al. 2007). Guidance instructed forecasters to identify an imminent or ongoing threat to human lives, which they further defined as a tornado "approaching a population center" (Warning Decision Training Division 2021). More recent modules carefully omit any mentions of population, however, as these markers differ significantly across tornado-prone regions (Warning Decision Training Division 2023a, b). Ongoing research is attempting to include other demographic information into the tornado emergency issuance process, and results have shown that knowledge of potentially vulnerable populations increased the likelihood that a forecaster would upgrade a tornado warning to a tornado emergency (Othling et al. 2023; Friedman et al. 2024). The extent to which demographic information is utilized operationally by NWS meteorologists is unclear, however, potentially varying between offices (Cook and Henson 2024).

While considerable research has been conducted into both tornado forecasting and tornado warning dissemination, tornado emergencies and the guidance for their issuance represent a gap in the literature. Studies of early usages of the "considerable" and "catastrophic" damage tags found low POD and high FAR for EF-3 tornado occurrence (Obermeier and Anderson 2014), though this was prior to the rollout of the WDTD guidance, which was developed in response to forecaster dissatisfaction with the nowcasting methods of the time (Harrison et al. 2014). However, given that the guidance supplied by the WDTD is noncompulsory and explicitly recommends that forecasters also use their own "judgement/experience" (Fig. 1), it remains to be seen how closely NWS meteorologists adhere to the provided thresholds and whether future iterations could more closely match forecasters' current practices.

The purpose of this article is to examine the dozens of tornado emergencies issued from 2014-2023 and determine whether each meets the current WDTD guidance for issuance. Additionally, we will explore whether any other characteristics, such as population density within the warning polygon and tornado path width and length, vary significantly between tornadoes that are warned with tornado emergencies versus those that are not. Any such differences may reveal information beyond VROT and STP that forecasters are considering when choosing to issue a tornado emergency. Lastly, we will evaluate how often tornado emergencies warn the most intense and deadly tornadoes and make recommendations for how future NWS training modules might better prepare forecasters to issue these types of warnings.

## 2. Methodology

While no official record of tornado emergencies is available, the Iowa Environmental Mesonet (IEM) does maintain an unofficial archive of these products (IEM 2025), which we manually verified for 2014-2023 by examining the warning text and identifying the use of the "catastrophic" phrasing, synonymous with tornado emergencies issued within the IBW framework. The dataset includes tornado emergencies issued as updates to existing warnings via severe weather statements. Tornado watch polygons and SPC convective outlooks were also accessed through their respective IEM archives.

To calculate VROT, the nearest NEXRAD site to each emergency polygon centroid was identified and radar data were imported into Gibson Ridge Level II Analyst software

(http://www.grlevelx.com/) from the NCEI radar archive (https://www.ncdc.noaa.gov/nexradinv/). In each instance, VROT was calculated from the lowest available elevation angle (usually 0.5°) for each scan in the 10 minutes prior to tornado emergency issuance. Care was taken not to record velocities associated with sidelobe contamination, which was present in a number of cases, by only examining velocity values within reflectivity ≥ 30 dBZ. Additionally, VROT was calculated independently by separate researchers, and the greater of the two values was recorded to ensure identification of the greatest possible value a forecaster could reasonably observe. The presence of a TDS was also noted, which we defined as a region of correlation coefficient values less than 0.8, co-located with a tornado vortex signature (TVS) and reflectivity ≥ 30 dBZ on the lowest elevation angle.

STP was computed using 13-km resolution analysis data from the NOAA Rapid Refresh (RAP) model (Benjamin et al. 2016) accessed from the NCEI Thematic Real-time Environmental Distributed Data Services (THREDDS) server (https://www.ncei.noaa.gov/thredds/catalog/model/rucrap.html). Data were selected from the top of the hour of tornado emergency issuance and masked to exclude convectively contaminated grid cells, following the methodology of Sessa and Trapp (2023). Fixed-layer and effective-layer STP were then computed using the Metpy package (May et al. 2022) for the unmasked grid cells within a 300-km-by-300-km grid centered on the centroid of the emergency polygon, and the maximum values within this search grid were recorded.[2] These particular formulations of STP were chosen as they closely mirror the versions used in studies referenced within training materials as being the basis for the WDTD thresholds (e.g., Smith et al. 2015), though no precise version is listed in the guidance.

A system for matching NCEI tornado tracks to tornado emergency polygons utilizing the Shapely python package (https://github.com/Toblerity/Shapely) was also developed. The characteristics of the intersecting tornado track with the highest EF rating were assigned to each tornado emergency. At this stage, county-level socioeconomic data from the Centers for Disease Control (CDC) Social Vulnerability Index (SVI; CDC 2018) were also joined to each tornado track. The tornado matching process was also repeated to assign all intersecting tornado tracks to each emergency polygon for the purpose of computing POD statistics. Lastly, maximum population density within each tornado emergency polygon was computed from NASA's Gridded Population of the World (GPWv4; CIESIN, 2017), which has a spatial resolution of approximately 1 km. Population densities were also computed within each base tornado warning polygon for days on which tornado emergencies were issued as a point of comparison.

[2] The use of maximum value within a 300-km grid is purposefully generous and was chosen to ensure that the highest value of STP a forecaster could potentially observe anywhere near the storm was recorded. This search radius is slightly less than that used by Thompson et al. (2012) to examine regional maxima for environmental parameters.

## 3. Results

*a. Spatial and Temporal Trends in Usage*

From 2014 to 2023, a total of 89 tornado emergencies were issued within the IBW framework. The majority of these were issued by WFOs within NWS Southern Region, specifically in the Southeast US (Fig. 2). This spatial maximum in tornado emergencies aligns with the maximum in tornadoes rated ≥ EF-3 over the same period. A number of tornado emergencies, particularly in the Southeast, were issued in sequence for a single storm, with 42% of tornado emergencies serving as follow-ups to previous emergencies. 48% of tornado emergencies were issued as severe weather statements (SVSs), which upgrade existing warnings to the "catastrophic" IBW tag. In other instances, tornado emergencies were nested inside of existing tornado warnings as standalone products (e.g., the Mayfield, KY tornado emergency in December 2021).

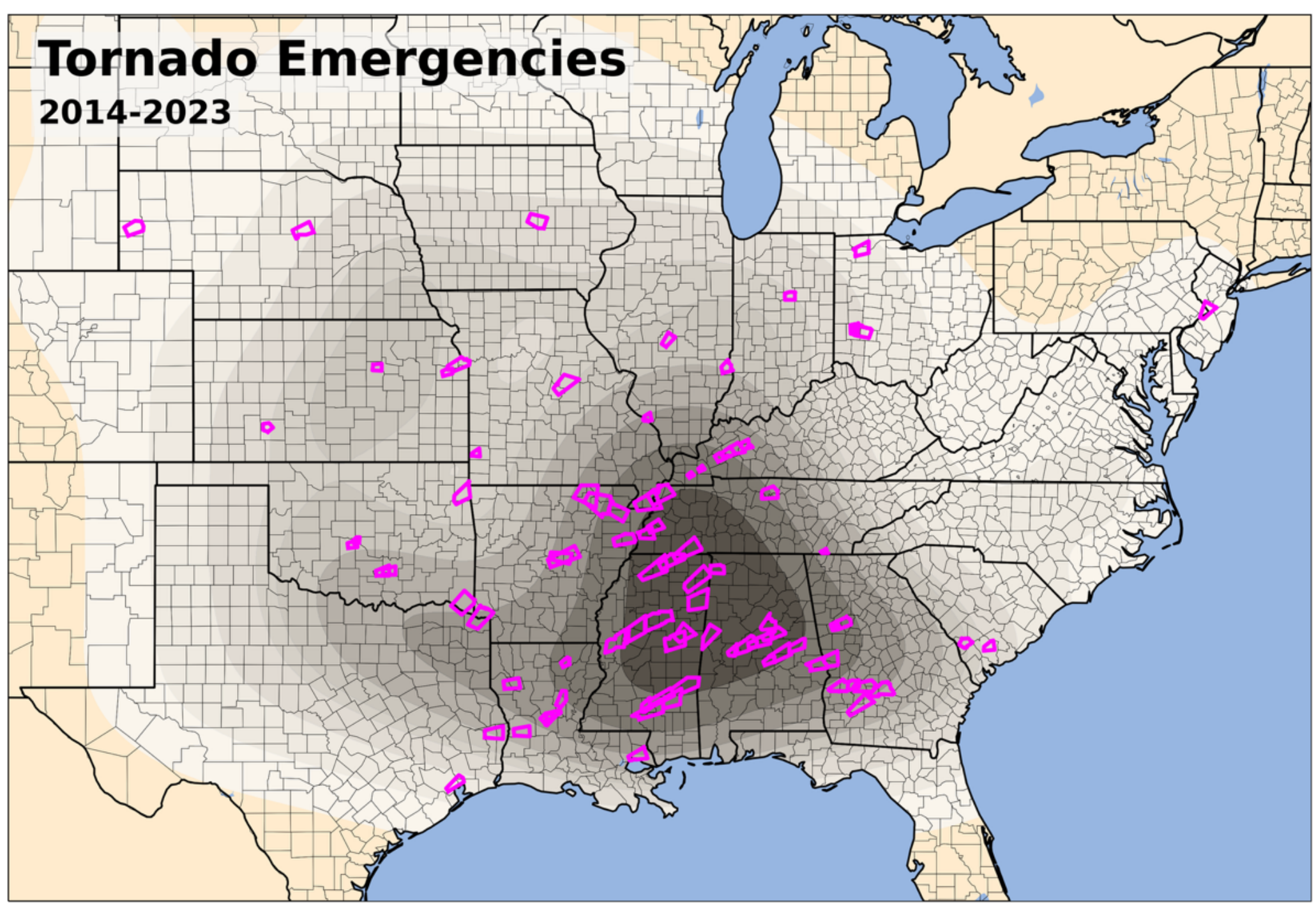

**Fig. 2.** Map of all 89 IBW tornado emergencies issued from 2014-2023. Filled contours are a kernel density estimate for all tornadoes rated ≥ EF-3 over the same period.

In addition to following other warnings, a significant number of tornado emergencies were also preceded by longer-fuse products. All emergencies were within some level of SPC convective outlook categorical risk: 83% were within an "enhanced" risk area or greater with 15% being issued within "high" risk areas. Only 3 tornado emergencies were issued within the lowest SPC risk category, "marginal." One of these emergencies was for an EF-3 tornado while the other two were for EF-1 tornadoes. Additionally, all but three tornado emergencies occurred within 25 km of an active tornado watch polygon and 27% occurred within a PDS tornado watch. Two of the emergencies not near tornado watches were false alarm events where no tornado occurred. Unsurprisingly given the number of associated watches, many tornado emergencies took place during relatively significant severe weather events. Using the Brooks et al. (2014) definition of a "tornado outbreak day," 30 or more EF-1+ tornadoes from 12 UTC to 11:59 UTC, we found that 29% of tornado emergencies occurred during outbreaks. Lowering the outbreak threshold to 15 or more EF-1+ tornadoes increases this percentage to 51%.

The number of IBW tornado emergencies issued each year has been generally increasing over the past decade (Fig. 3a), potentially due in part to an increased familiarity with the product among NWS WFOs. Nevertheless, tornado emergencies still only account for a small percentage of the total number of tornado warnings issued each year by the NWS, less than 1% (Fig. 3b). This is to be expected as tornado emergencies target particularly violent, deadly tornadoes, which represent only a fragment of all tornadoes each year. Examining the EF ratings of the strongest tornadoes contained within each emergency polygon shows that the majority of emergencies, 70%, were issued for EF-3 and EF-4 tornadoes (Fig. 4), though some weaker tornadoes were also warned. The 3 false alarm tornado emergencies, noted by the "N/A" bar in Figure 3, as well as other performance metrics are discussed in more detail in Section 3b.

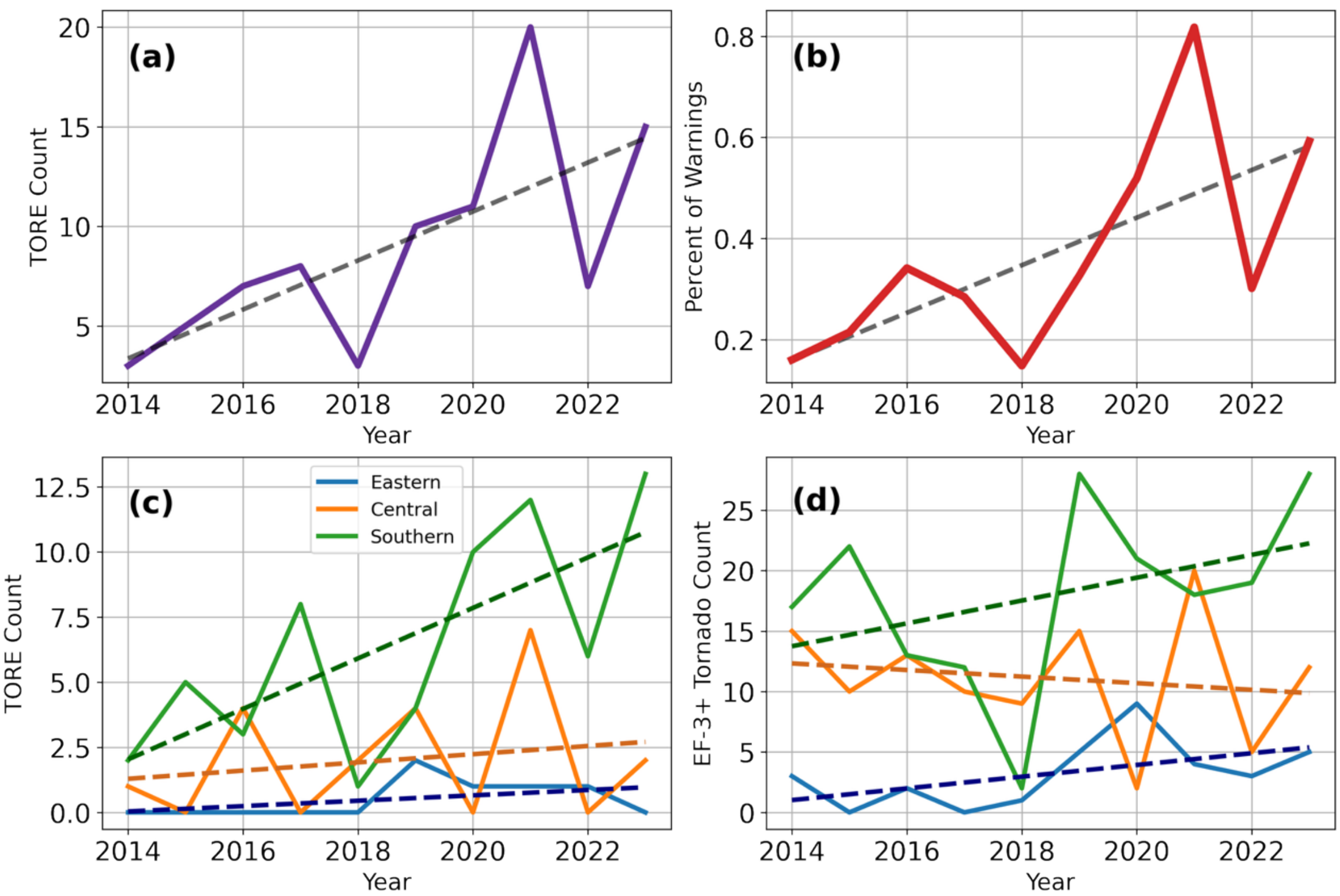

**Fig. 3.** Yearly trends in (a) nationwide tornado emergency (TORE) counts, (b) nationwide tornado emergency counts as a percentage of all tornado warnings, (c) tornado emergency counts separated by NWS region, and (d) ≥ EF-3 counts, also separated by NWS region. Note that tornado emergencies have only been issued in Eastern, Central, and Southern Regions. In each panel, dotted lines represent best fit lines through the data. Excluding the first three years of the period, when IBWs were not fully implemented, does not meaningfully impact the trends shown.

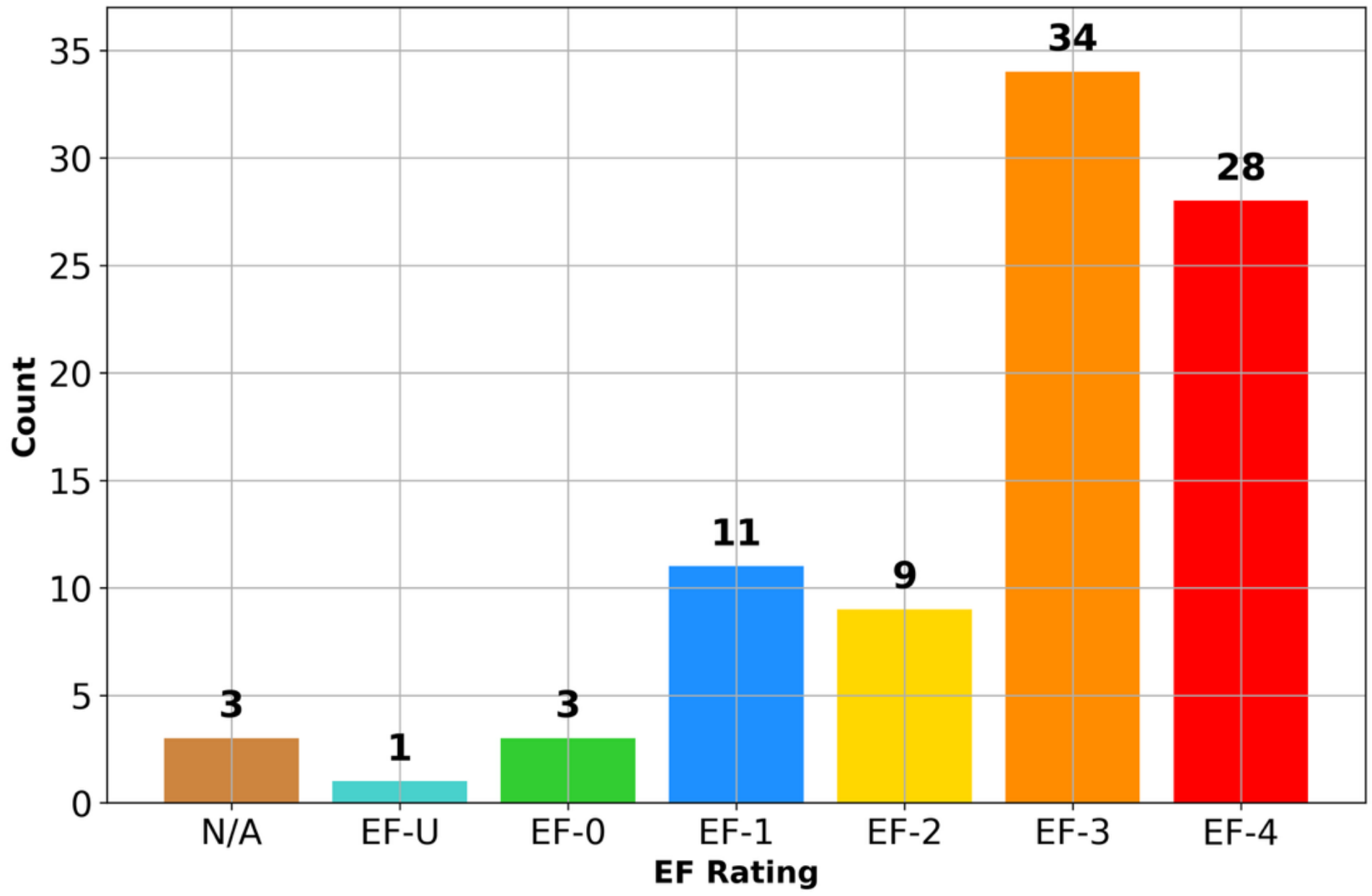

**Fig. 4.** Distribution of tornado emergencies by maximum EF rating, with the N/A bar denoting false alarms (i.e., emergencies for nontornadic storms).

The majority of the increase in tornado emergency issuance comes specifically from NWS Southern Region (Fig. 3c), as expected given that the most emergencies were issued by their WFOs. The increase in tornado emergencies in NWS Southern Region is only partially attributable to an increase in particularly intense tornadoes (Fig. 3d), with 24% of the variance in annual tornado emergency counts explainable by the number of EF-3+ tornadoes in those years ($R^2 = 0.24$; Fig. 5). Thus, we hypothesize that factors beyond simply the occurrence of intense tornadoes may have driven an increase in the usage of tornado emergencies. Future work will be needed to address whether evolving policies and philosophies surrounding tornado emergencies at the regional level and among individual forecast offices may influence decision processes for tornado emergencies. The following analysis will attempt to elicit how application of the WDTD guidance may vary by region.

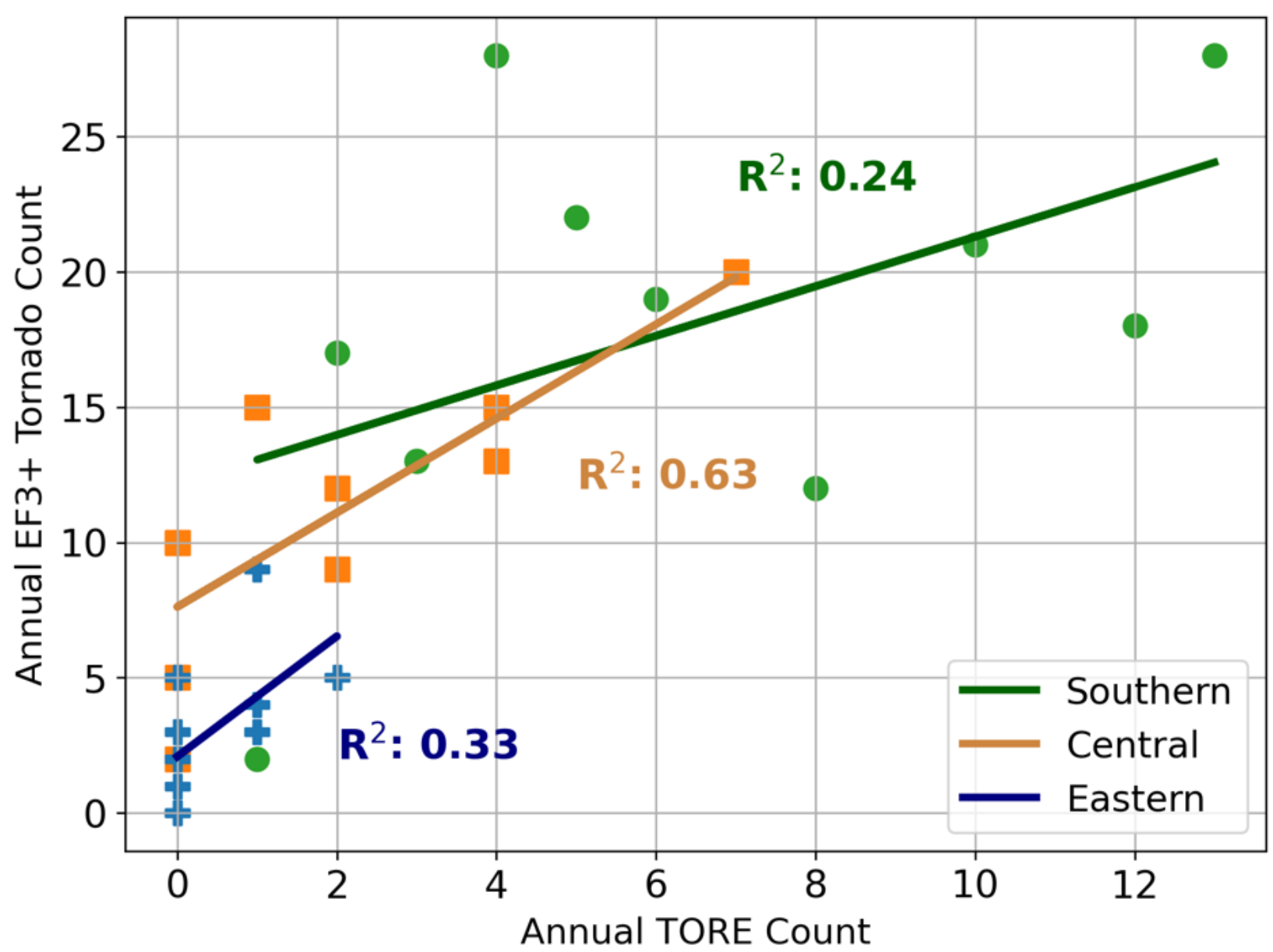

**Fig. 5.** Scatterplot showing the relationship between annual tornado emergency counts and the number of EF-3+ tornadoes in each year, separated by NWS region. Linear regressions for each region and their corresponding $R^2$ values are also shown. Note that some data points are obscured by other points from the same region.

*b. False Alarms and Performance Metrics*

In order to gauge the performance of tornado emergencies, it is first necessary to define the purpose of these warnings in a quantifiable way. We therefore compute FAR, POD, and critical success index (CSI; Doswell et al. 1990) following the formulations in Brooks (2004b). Applying the same metric used for all tornado warnings, i.e., the presence of a tornado, shows that tornado emergencies have an impressively low FAR, 0.03, as compared to the FAR for tornado warnings as a whole, 0.69 (Keylon 2025). This is significant as one of the motivating factors for the creation of the IBW program was to lower the perceived number of false alarms (NOAA 2011). Given that a criterion for tornado emergency issuance is confirmation of a tornado, a low FAR is to be expected. In total, there were 3 tornado emergency false alarms, all of which were issued for the same nontornadic thunderstorm in northern Arkansas on 16 April 2022. As reported in the following days, severe straight-line winds and poor visibility likely contributed to false real-time reports of tornado damage (Navarro 2022). Our analysis also found that both VROT and STP were well below WDTD thresholds (see Section 3c), though VROT analysis was hampered by sidelobe contamination and the distance to the nearest radar (>150 km).

Of course, the purpose of a tornado emergency is explicitly not to warn every tornado, but specifically situations where "catastrophic" damage is likely (NWS 2023). This description can be interpreted as referring to either tornadoes ≥ EF-4 strength (Obermeier and Anderson 2014) or tornadoes ≥ EF-3 strength, based upon VROT and STP values from Smith et al. (2015). Tornado emergencies warned 23% of all EF-3+ tornadoes and 55% of all EF-4 tornadoes (Fig. 6a), with no EF-5 tornadoes occurring in this period. At these thresholds the FAR is 0.3 and 0.69 respectively. This is still a remarkable achievement on the part of forecasters and researchers, as predicting tornado intensity in addition to tornado occurrence is a substantially more difficult task. It should also be noted that there are inherent issues with using EF rating to gauge tornado intensity (e.g., Lyza et al. 2024).

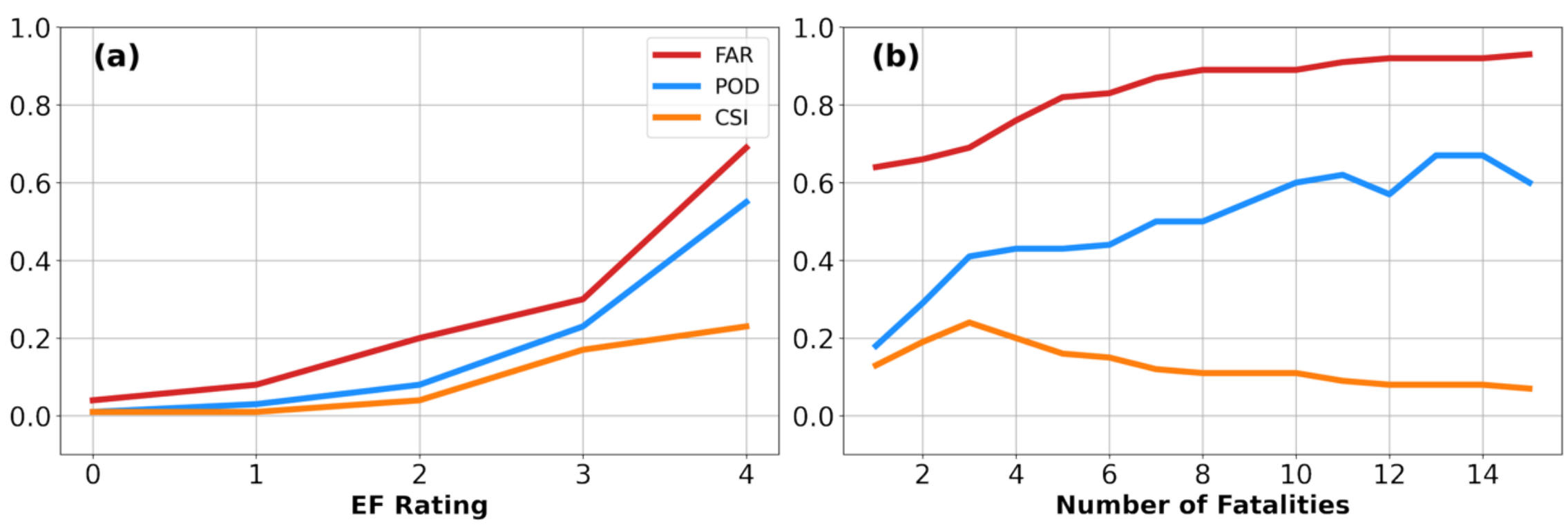

**Fig. 6.** Plots of tornado emergency POD, FAR, and CSI by (a) EF rating and (b) number of fatalities. Each point on the x axis corresponds to values of these skill metrics for that threshold or greater. For example, the FAR shown at EF-3 indicates the fraction of tornado emergencies issued for either tornadoes < EF-3 intensity or for nontornadic storms.

While tornado emergencies have only warned 18% of all fatal tornadoes, they have warned many of the deadliest tornadoes, including 41% of all tornadoes with 3 or more fatalities and 60% of tornadoes with 10 or more fatalities (Fig. 6b). 64% of emergencies were for nonfatal tornadoes, though it is impossible to separate the effects that tornado intensity, socioeconomic factors, and the use of enhanced wording (i.e., the tornado emergency itself) all have on the likelihood of fatalities.

*c. Issuance Criteria*

1) VROT

In total, 61.8% of tornado emergencies featured VROT greater than or equal to the recommended 70 kt threshold supplied by the WDTD in the 10 minutes prior to tornado emergency issuance (Fig. 7a). Breaking down VROT values by NWS region reveals no statistically significant differences as determined using a non-parametric Kruskal-Wallis test, which tests whether samples originate from the same distribution (Kruskal and Wallis 1952). The median 10-minute maximum VROT is approximately the same for each region and all medians fall above the 70 kt WDTD threshold (Fig. 7b). These results indicate that a majority of tornado emergencies are issued with VROT exceeding the WDTD criteria across all three NWS regions.

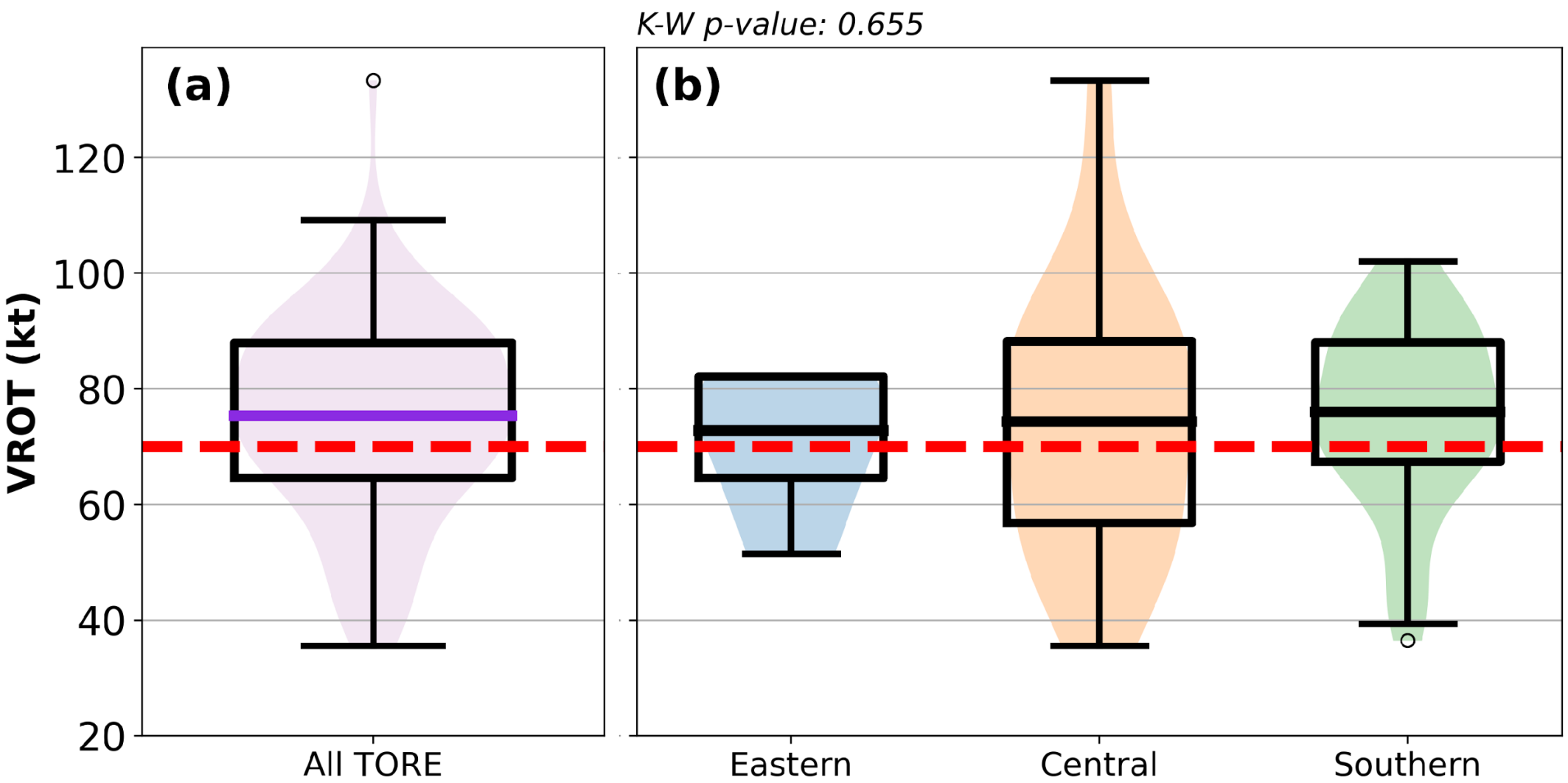

**Fig. 7.** Boxplots of maximum rotational velocity over the 10 minutes prior to tornado emergency issuance for each tornado emergency, (a) for all emergencies nationwide and (b) separated by NWS region. Dashed red lines denote the WDTD threshold of 70 kt in both panels. The *p*-value from a Kruskal-Wallis test testing for statistically significant differences across the 3 regions is listed above panel (b).

While the advertised VROT threshold is 70 kt, training materials do suggest lowering this value by 5-10 kt in instances where significant debris is present (Warning Decision Training Division 2023a). If we apply a 60 kt threshold to any cases with a TDS present at the last radar scan prior to emergency issuance, a total of 71 cases or 79.8% of all emergencies now meet the VROT threshold. Further research would be needed to determine whether forecasters are applying this correction or simply not utilizing the WDTD threshold.

Other factors may influence VROT values and the appearance of a tornado or mesocyclone on radar, such as the distance to the nearest NEXRAD site. At long ranges, TVSs may appear weaker due to under sampling and the effect of random beam positioning relative to the azimuth of maximum velocities (Burgess et al. 1993; Wood and Brown 1997), something that is noted in WDTD guidance (e.g., Warning Decision Training Division 2023a). However, we computed the distance to the nearest radar for all tornadoes in the 2014-2023 period and found that those warned by tornado emergencies were not significantly closer to radar sites as compared to all other tornadoes when evaluated using a two-sided Mann-Whitney $U$ test ($p = 0.27$).

Another complicating factor is tornado width. Studies have shown that wider tornadoes tend to be associated with wider mesocyclones (Trapp et al. 2017; Sessa and Trapp 2020), both of which would lead to a larger rotation signature on radar. Across every EF category, the median width for tornadoes warned with emergencies was higher than those that did not receive these enhanced warnings (Fig. 8). For EF-2, EF-3, and EF-4 tornadoes, the difference is statistically significant per a Mann-Whitney $U$ test ($p = 0.04$, $p \ll 0.001$, and $p = 0.002$ respectively). Whether forecasters are explicitly evaluating mesocyclone width on radar or whether they are considering tornado width unconsciously remains unclear. It may be prudent for tornado emergencies to target the widest tornadoes, however. Wider tornadoes would take longer to pass over a point, assuming constant translational velocity, impacting structures with intense winds for a longer period and potentially resulting in more severe damage (Brooks 2004a). A wider tornado would also impact more structures than a narrower tornado, assuming identical path lengths. This is evident in the tornado record, as the widest tornadoes tend to account for a disproportionate percentage of fatalities (Garner et al. 2021).

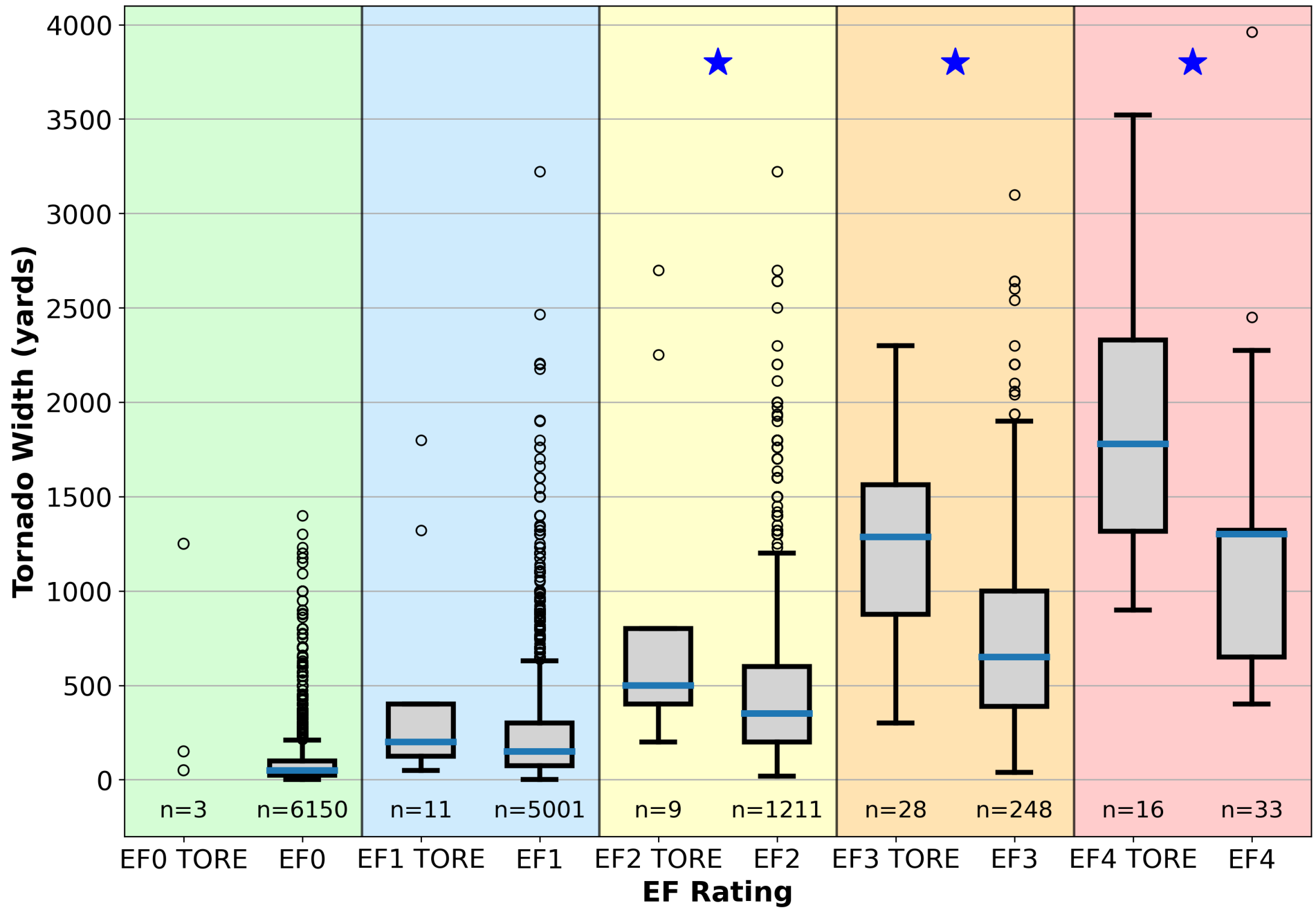

**Fig. 8.** Comparison of tornado width across EF ratings (2014-2023), subdivided by tornadoes with and without tornado emergencies. As there were only 3 EF-0 tornado emergency cases, these are represented by individual points rather than a boxplot. Blue stars over a pair of boxplots indicate a statistically significant difference in distribution, as determined by a Mann-Whitney *U* test, at the 0.05 level.

Likewise, tornadoes with longer path lengths are also more likely to be warned with tornado emergencies than tornadoes with shorter paths and, for EF-3 and EF-4 tornadoes specifically, this relationship is statistically significant ($p << 0$ and $p = 0.005$). This is notable as some public-facing tornado emergency descriptions state that storms warned with tornado emergencies are expected to be long-lived (e.g., NWS 2016). EF rating is strongly tied to tornado path length, both because tornadoes with longer paths tend to be more intense and because there are more opportunities for a long-track tornado to strike damage indicators capable of giving a higher EF rating (Straka and Kanak 2022; Straka et al. 2024) A lack of viable damage indicators could place a tornado into a lower EF category than its true wind speed (Lyza et al. 2024), however, potentially impacting the results shown here.

2) SIGNIFICANT TORNADO PARAMETER

Of the 89 selected tornado emergencies, 75 occurred at times for which RAP analysis data was available and at least 10% of grid cells were not convectively contaminated. In total, only 10.7% (12.8%) of these tornado emergencies met the WDTD threshold using fixed-layer (effective-layer) STP. The median value of fixed-layer STP for the examined emergencies was 4.1 (Fig. 9a) while the median of effective-layer STP was 3.3 (Fig. 9c).[3] For NWS Southern Region (Fig. 9b), both STP values are significantly lower than those in NWS Central Region ($p = 0.017$ and $p = 0.05$). Medians for each region are below the STP threshold of 6, though for NWS Central Region, it is closer to the threshold than the other two regions for both STP variants (Figs. 9b and 9d).

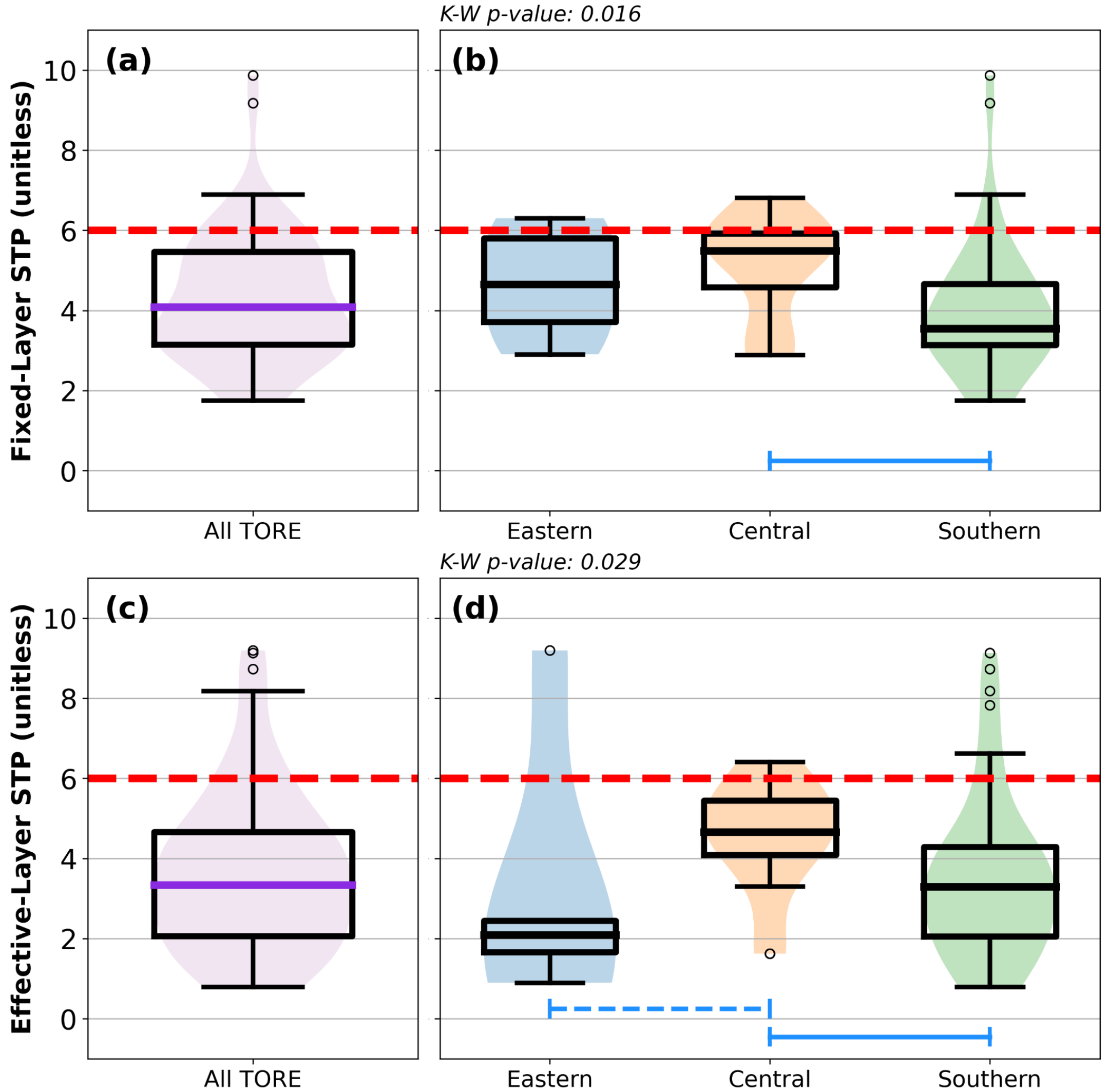

**Fig. 9.** As in Fig. 7 but for (a, b) fixed-layer STP and (c, d) effective-layer STP, with the red line denoting the WDTD threshold of 6. The dashed (solid) blue lines in panels b and d denote a statistically significant difference with $p$-value $\leq 0.1$ ($p$-value $\leq 0.05$) between the two connected groups as determined from a post-hoc Dunn's test.

In addition to computing STP for tornado emergencies, we also calculated this parameter for all particularly intense tornadoes ($\geq$ EF-3) from 2014 to 2023. Comparing fixed-layer STP for tornadoes with and without tornado emergencies (not shown), no statistically significant differences for either EF-3 ($p$ = 0.71) or EF-4 tornadoes ($p$ = 0.63) are found. If the STP criterion were being used, one would expect to see particularly intense tornadoes with high STP disproportionately falling into the tornado emergency category. This provides additional evidence that STP may not be considered by forecasters. The especially low median STP in NWS Southern Region may stem from known limitations of STP in the Southeast U.S. (Sherburn and Parker 2014; Brown et al. 2021). Our analysis of STP also revealed that only 11% of *all* EF-3+ tornadoes in the study period meet the WDTD threshold of 6, highlighting the potential limitations of this parameter in identifying intense tornadoes. Additional research is needed to assess how the WDTD STP criterion may be impacting forecasters decisions regarding the issuance of tornado emergencies, and whether this guidance threshold should be adjusted or removed.

[3] While the analyses shown here utilize fixed and effective-layer STP, forecasters may be using different versions of STP which could yield values that meet the 6 STP threshold in cases where these formulations do not. Thus, there may be some instances where forecasters are following the WDTD STP criteria that are not apparent from this analysis.

Examining both VROT and fixed-layer STP reveals that two tornado emergencies from 2014 to 2023 met both criteria as suggested by the WDTD, both for EF-3 tornadoes (Fig. 10a). Effective-layer STP yields four emergencies meeting both thresholds, issued for one EF-4 and three EF-3 tornadoes (Fig. 10b). Tornado emergencies issued with especially low VROT and STP tended to be less severe, with all 3 false alarm tornado emergencies falling into this category along with several EF-1 tornadoes. That said, several tornadoes rated ≥ EF-2 also fell below both WDTD thresholds. Given that so few cases meet both the VROT and STP thresholds, ongoing research is examining forecasters' decision-making processes when issuing TOREs, which may help identify best practices to inform future guidance.

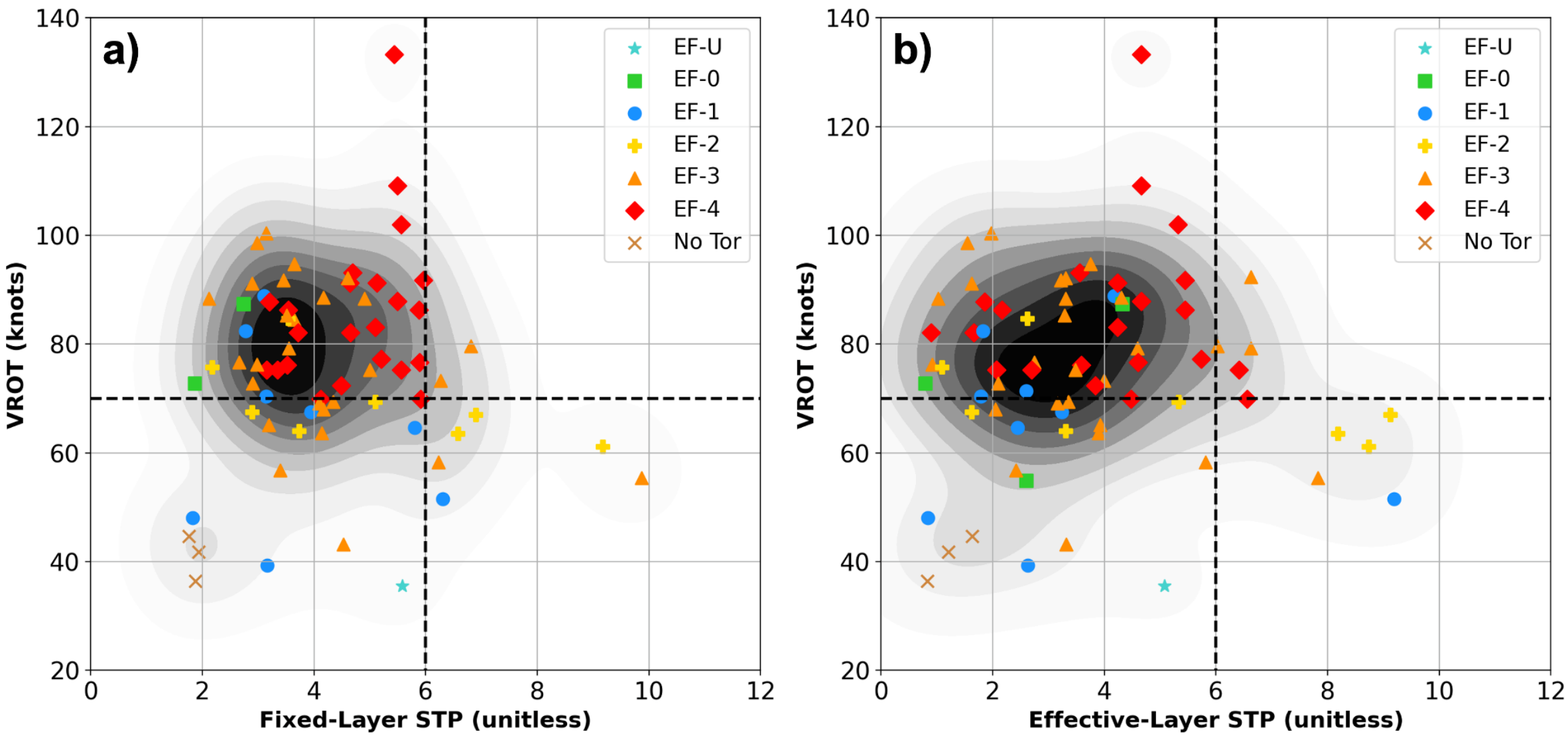

**Fig. 10.** Scatterplot of the maximum recorded VROT in the 10 minutes preceding emergency issuance and maximum (a) fixed-layer STP and (b) effective-layer STP from the top of the hour of tornado emergency issuance for each tornado emergency, with each point also denoting EF rating. WDTD thresholds are indicated by dashed lines, with the top right quadrants containing emergencies which met both criteria. Grayscale shading shows the kernel density estimate of the points. Note that only the subset of 75 tornado emergencies with STP values available to analyze are shown here.

3) DEBRIS SIGNATURES

While not one of the two primary criteria in the WDTD guidance, forecasters are suggested to seek out either spotter or radar confirmation of a tornado, with the latter coming from the presence of a TDS. Of the 89 tornado emergencies examined, 91% had a TDS in the last scan prior to emergency issuance. Of the 8 emergencies where a TDS was absent, 3 were false alarms, that is, cases where no tornado was present. However, this does not indicate that tornadoes were not confirmed at the time of emergency issuance in these remaining 5 cases. The emergency text notes that spotters confirmed a tornado visually in 4 cases while the fifth may have had a TDS present before the last scan prior to tornado emergency issuance.

4) POPULATION DENSITY

Though no longer mentioned in training materials, there is evidence that some forecasters likely do consider population density when issuing tornado emergencies, specifically targeting tornadoes impacting population centers. In NWS Central Region, the median maximum population density contained within tornado emergency polygons is statistically significantly higher than for all other tornado warnings issued by the same WFOs on the same days (Fig. 11). There is little difference in the medians for NWS Southern Region, perhaps highlighting a different approach to tornado emergency issuance by those WFOs. Leadership from the Jackson, MS WFO, for example, have stated that their policy is not to consider population density due to the large number of rural residents in their CWA (Cook and Henson 2024). The influences of debris sources within urban areas on TDS appearance (e.g., Cross et al. 2023) may also affect the issuance of emergencies and is a topic in need of additional research especially as debris signatures become increasingly important in warning operations.

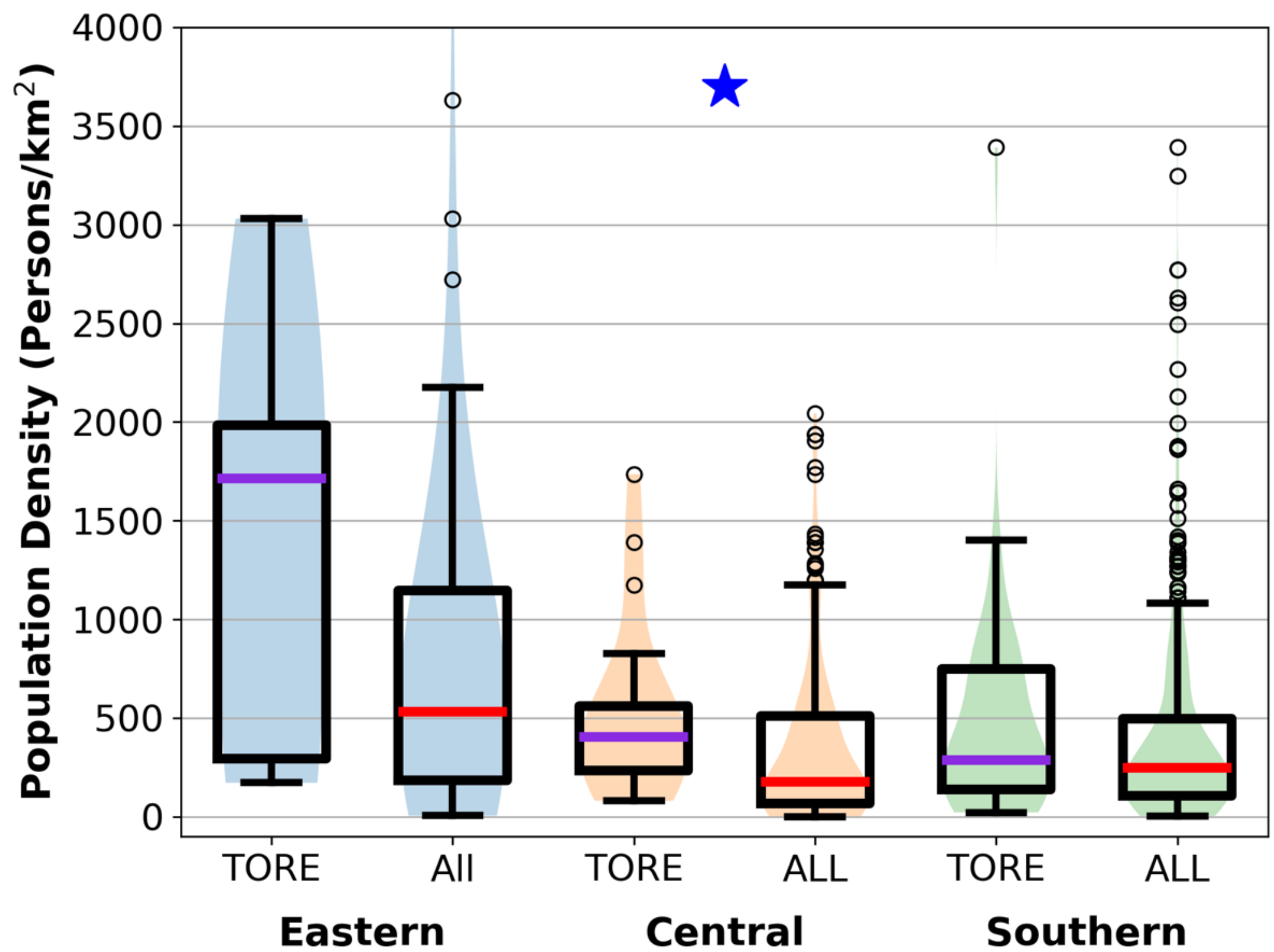

**Fig. 11.** Boxplots of maximum population density within both tornado emergency polygons and all tornado warning polygons issued on the same days and by the same WFOs as the emergencies. As in Fig. 7, data are subdivided by NWS region. A blue star over a pair of boxplots indicates a statistically significant difference in distribution, as determined by a Mann-Whitney *U* test, at the 0.05 level. Note that two outlier values from the "All" portion of the Eastern Region plot (6500 and 8000 person km$^{-2}$) are not shown.

5) SOCIOECONOMIC FACTORS

Although researchers have suggested avenues for forecasters to consider socioeconomic factors when issuing tornado emergencies (Othling et al. 2023; Friedman et al. 2024), it is not clear whether they have begun to do so operationally. Most variables in the CDC SVI dataset vary considerably between the 3 EF-0 tornadoes warned with emergencies and the remaining EF-0 tornadoes. This includes the percentage of residents in mobile homes, the percentage of those without access to a vehicle, and the percentage with income below the poverty line, all of which were higher for EF-0 tornadoes which received tornado emergencies. All three of these tornadoes occurred in the Southeast US and in fairly rural areas. Notably, two of the three had VROT values above the 70 kt threshold. It is not known whether the socioeconomic vulnerability of these areas led forecasters to upgrade to tornado emergencies. It is also possible that the winds observed aloft on radar simply did not translate to the surface in these cases as all EF-0 tornadoes were > 50 km from the nearest radar site.

## 4. Conclusions and Recommendations

Given the disparate impacts of weak versus intense tornadoes, it logically follows that the wording of tornado warnings should also vary in severity, especially with advances in nowcasting tornado intensity. The formalization of the tornado emergency has allowed the NWS to begin warning not just the occurrence of tornadoes but also the expected impacts. However, relatively little research has explored the outcomes of tornado emergencies since their formalization, specifically how often forecasters utilize the empirical guidelines for issuance and how well these warnings actually discriminate between less impactful tornadoes and particularly rare, catastrophic events.

Our analysis of the first decade of formalized tornado emergencies has produced the following conclusions:

- The FAR for tornado emergencies is much lower than that of all tornado warnings, including when validating by EF-3 tornado occurrence, with only 30% of tornado emergencies issued for < EF-3 tornadoes. 55% of all EF-4 tornadoes and 41% of tornadoes with ≥ 3 fatalities were warned by tornado emergencies.
- Upgrades to tornado emergencies typically occur when high values of VROT are present. The majority of cases examined (61.8%) had VROT greater than or equal to the 70 kt WDTD threshold in the 10 minutes prior to emergency issuance.
- The specific STP criterion developed by the NWS is not met in a vast majority of cases, as only 10.7% (12.8%) of tornado emergencies were issued for environments with fixed-layer (effective-layer) STP ≥ 6. Notably, only 11% of *all* EF-3+ tornadoes during this period occurred in environments with fixed-layer STP ≥ 6.
- Population density within tornado emergencies is significantly higher than other tornado warnings in NWS Central Region, while this difference is not observed in NWS Southern Region.

Based on these conclusions, we make the following recommendations for future research:

- The WDTD STP threshold should be further studied in operational contexts. Previous studies have emphasized relatively poor performance of standard thresholds of this parameter, particularly in the Southeast (Sherburn and Parker 2014; Brown et al. 2021).
- New metrics for nowcasting tornado intensity, such as updraft characteristics (Marion et al. 2019; French and Kingfield 2021; Wolff et al. 2025), pre-tornadic mesocyclone width (Sessa and Trapp 2020; Sessa and Trapp 2023), and tornado debris signature height (Bodine et al. 2013; Van Den Broeke and Jauernic 2014; Cross et al. 2023), should continue to be developed and eventually evaluated for inclusion in WDTD guidelines.[4] Notably, the latter two tools would not require forecasters to shift focus away from radar analysis during warning operations.

- Further research should examine whether the inclusion of socioeconomic factors, such as mobile home population, households without access to vehicles, etc. could further improve the ability of tornado emergencies to discriminate between deadly and non-deadly tornadoes when used in concert with radar-based analysis (e.g., Othling et al. 2023; Friedman et al. 2024).
- Forecasters with experience issuing tornado emergencies should be interviewed to better understand what information they considered during the warning process and how they self-evaluated the success of their warnings, building upon the insights into one office's procedures provided by Cook and Henson (2024) and initial evaluations of forecaster satisfaction with IBWs from Harrison et al. (2014).

Future research should focus on identifying and evaluating new methods for nowcasting tornado intensity while also improving understanding of the current practices of forecasters. As our results show, the validation of a tool in research settings and introduction to forecasters through training modules alone does not guarantee its successful implementation. Our research group has already begun to conduct semi-structured interviews with NWS meteorologists who have issued tornado emergencies, with a focus on individual events, which we hope will illuminate forecaster decision-making processes in greater detail. Additionally, the analysis conducted here could be extended to warnings with the "considerable" damage tag and their associated WDTD guidelines. Although some research has examined public response to impact-based messaging (e.g., Casteel 2016), further work may elucidate new results after a decade of exposure to such warnings. In particular, public perception of tornado emergencies could differ considerably from their intended purpose and may unintentionally lower the perceived risk associated with untagged tornado warnings. The interpretation of warnings across different languages also remains an open and important question, particularly given documented inconsistencies in warning responses within bilingual communities (Trujillo-Falcón et al. 2021; 2024). Insights into these topics could be gleaned from nationwide surveys and targeted interviews in communities which have experienced impactful tornadoes.

[4] Although these tools are not included as official criteria for tornado emergency issuance, many are already covered in depth in WDTD training modules.

The goal of warning criteria is to provide forecasters, particularly those less familiar with warning operations, with a framework for issuing these rare warnings. Though there will always be the need for flexibility, an ideal set of guidelines should match the best practices of experienced forecasters as closely as possible to minimize the need for on-the-job learning after the completion of training courses. By working to better understand the current de facto tornado emergency criteria employed by NWS meteorologists and suggesting improvements where necessary, a more useful set of guidelines can be developed, hopefully leading to even more consistent and accurate warnings in future decades.

*Acknowledgments.*

Funding for the publication of this manuscript was provided by start-up funds awarded to J.T.F. from the Department of Climate, Meteorology, & Atmospheric Sciences at the University of Illinois Urbana-Champaign. We greatly appreciate the work of Daryl Herzmann (IEM) to archive TORE data and for answering our questions. This manuscript was greatly improved by comments from Jeff Waldstreicher and 3 anonymous reviewers as well as conversations with Andy Hatzos (NWS Wilmington, OH) and Lexy Elizalde-Garcia (NWS WDTD). Our thanks to the NWS and WDTD for their tireless work to inform the public of hazardous weather.

*Data Availability Statement.*

All relevant Python scripts and datasets are available on GitHub (https://github.com/EWolffWX/BAMS_TORE).